\documentclass[%
 reprint,
 amsmath,amssymb,
 aps,
]{revtex4-2}

\usepackage{graphicx}
\usepackage{dcolumn}
\usepackage{bm}
\usepackage{placeins}

\begin{document}


\title{Diode Effect May Assist Finding Proper Superconductivity Mechanism\\ in Copper Oxides}%

\author{Armen Gulian}
 \email{Corresponding author: gulian@chapman.edu}
\author{Serafim Teknowijoyo}%
\author{Vahan Nikoghosyan}
\altaffiliation[Also at ]{Institute for Physical Research, National Academy of Sciences, Ashtarak, Republic of Armenia.}

\affiliation{%
 Laboratory of Advanced Quantum Materials and Devices,\\Institute for Quantum Studies, Chapman University, USA 
}%

 \homepage{http://www.advanced-qmd.squarespace.com}


\date{\today}

\begin{abstract}
We present measurements demonstrating that copper-oxide high-temperature
superconductors can exhibit broken time-reversal symmetry in the absence of
external magnetic fields. Using $Tl_{2}Ba_{2}CaCu_{2}O_{8}$ microbridges, we
observe a pronounced superconducting diode effect at 100 K under strictly
zero-field conditions. This nonreciprocal response remains unchanged in
magnetic fields up to $\pm 100 Oe$. Our results are consistent with recent
reports of zero-field diode behavior in $Bi_{2}Sr_{2}CaCu_{2}O_{8+{\delta} }$
and together indicate that time-reversal symmetry breaking may be an
intrinsic property of the cuprate superconducting state. These findings
significantly constrain theoretical models of high-temperature
superconductivity that rely on time-reversal-symmetric mechanisms.
\end{abstract}

\maketitle


\section{\label{sec:level1}Introduction}

High-temperature superconductivity (HTSC) in copper oxides has been known
for nearly forty years, yet the mechanism behind this phenomenon remains
unresolved. Since the time of Sir Neville Mott, it has often been said that
there are as many theories as there are theorists. Despite decades of
progress, the number of proposed or modified theoretical models has not
decreased significantly, although experimental results have helped identify
several leading candidates. However, their validation is critically dependent on targeted experimental tests.

A particularly informative development has emerged recently from research on the superconducting diode effect. This effect, reminiscent of the
semiconductor diode, manifests itself as the suppression of one unipolar branch of
an alternating current flowing through a superconducting system, as
illustrated in Fig. 1.

\begin{figure}
\includegraphics[width=0.9\linewidth]{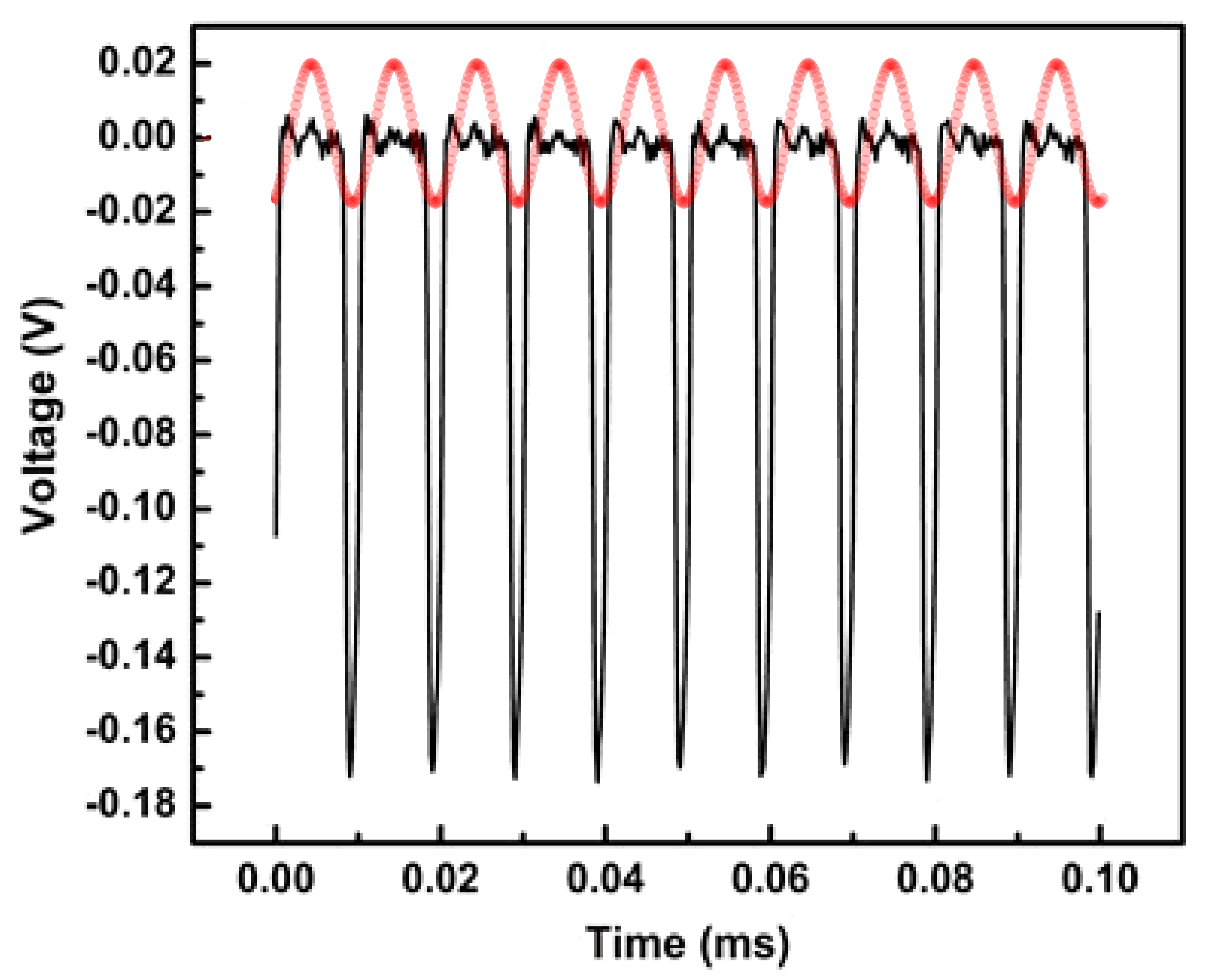}
\caption{Superconducting diode effect in presence of external magnetic
field: the voltage down-spikes in response to applied sinusoidal current (shown in arb. units, circles). These data were obtained in our laboratory with $Nb_{3}Sn$ bridge at $2 K$; the magnetic field is applied
perpendicular to the film surface, more details can be found in \cite{ref1}.}
\label{Figure1}
\end{figure}
First clearly demonstrated in $Nb/V/Ta$ superlattices \cite{ref2} through
the observation of nonreciprocal critical currents, the effect has since
been reported in a wide range of systems, including artificially structured
materials (such as those with conformal holes \cite{ref3}), Josephson
junctions, and microbridges (see the recent review \cite{ref4}).

A characteristic feature of systems exhibiting this effect is the
simultaneous breaking of time-reversal symmetry (TRS) and inversion symmetry
(IS). This combination is readily identified within the Ginzburg-Landau
framework \cite{ref5}. Typically, TRS breaking is associated with the presence
of a magnetic field, while IS may be absent due to either the crystal
structure or environmental factors. In most experimental settings the
magnetic field is externally applied, although exceptions exist, such as
chiral order parameters in triplet superconductors \cite{ref6} or magnetic
inclusions.

Although initial studies focused on conventional superconductors,
diode-effect research has now extended to the cuprate HTSCs \cite{ref7}. A
particularly notable result is that the diode behavior in $%
Bi_{2}Sr_{2}CaCu_{2}O_{8+{\delta} }$ flakes appears in the absence of any
externally applied magnetic field, with the effect observed up to $72 K$ for
samples with $T_{c}^{onset}=93.6 K$, and $T_{c}^{zero}=75.5 K$. The appearance
of a diode effect under strictly zero-field conditions requires careful
interpretation. The authors of Ref.\cite{ref7} attributed this to TRS
breaking caused by internal electronic effects - specifically, loop-current
states associated with the pseudogap phase in Varma's model \cite{ref8}.

In the present paper we extend the investigation of Ref.\cite{ref7} to
lithographically patterned microbridges fabricated from the related cuprate
superconductor $Tl_{2}Ba_{2}CaCu_{2}O_{8}$, with measurements performed at $%
100 K$. Section 2 describes the physical properties of the films and bridges.
We show that this simple microbridge structure exhibits a diode effect at $%
100 K$ under strictly zero magnetic field. Furthermore, measurements carried
out under fields of $\pm 100Oe$ show no deviation from the zero-field
behavior. Since our results are consistent with those reported in
Ref.\cite{ref7} and more recently in Ref.\cite{ref9}, they are likely to be further corroborated, and significantly
constrain theoretical models of HTSC in cuprates that do not involve TRS
breaking. Section 3 discusses the classes of theories compatible or
incompatible with these observations. Section 4 summarizes the implications
of our findings and outlines possible directions for future work. Section 5
describes the methods employed.

\bigskip
\section{\label{sec:level1}Experimental details}

The films of $Tl_{2}Ba_{2}CaCu_{2}O_{8}$ were deposited as described in \cite%
{ref10} on $LaAlO_{3} $ substrates using $Ba-Ca-Cu-O$ precursors. Bridges
were defined by wet photolithography using iron chloride as an etchant. The
resistive and magnetic properties of the films were measured with a PPMS
DynaCool (Quantum Design). Bridge transport measurements used a Keithley
6221 current source and a Keithley 2182A nanovoltmeter for low-voltage
measurements, and a Keithley DMM 2000 for higher-voltage acquisitions. All
instruments were connected externally to the DynaCool cryostat. In DC mode,
the $I+$ branch was swept from zero to its maximum value; the polarity of
the current leads was then reversed to obtain the corresponding $I-$ branch.

\bigskip
\section{\label{sec:level1}Results}

Our principal finding - the appearance of a diode effect at $100 K$ - is
presented in Fig. 2. 
\begin{figure}
\includegraphics[width=0.9\linewidth]{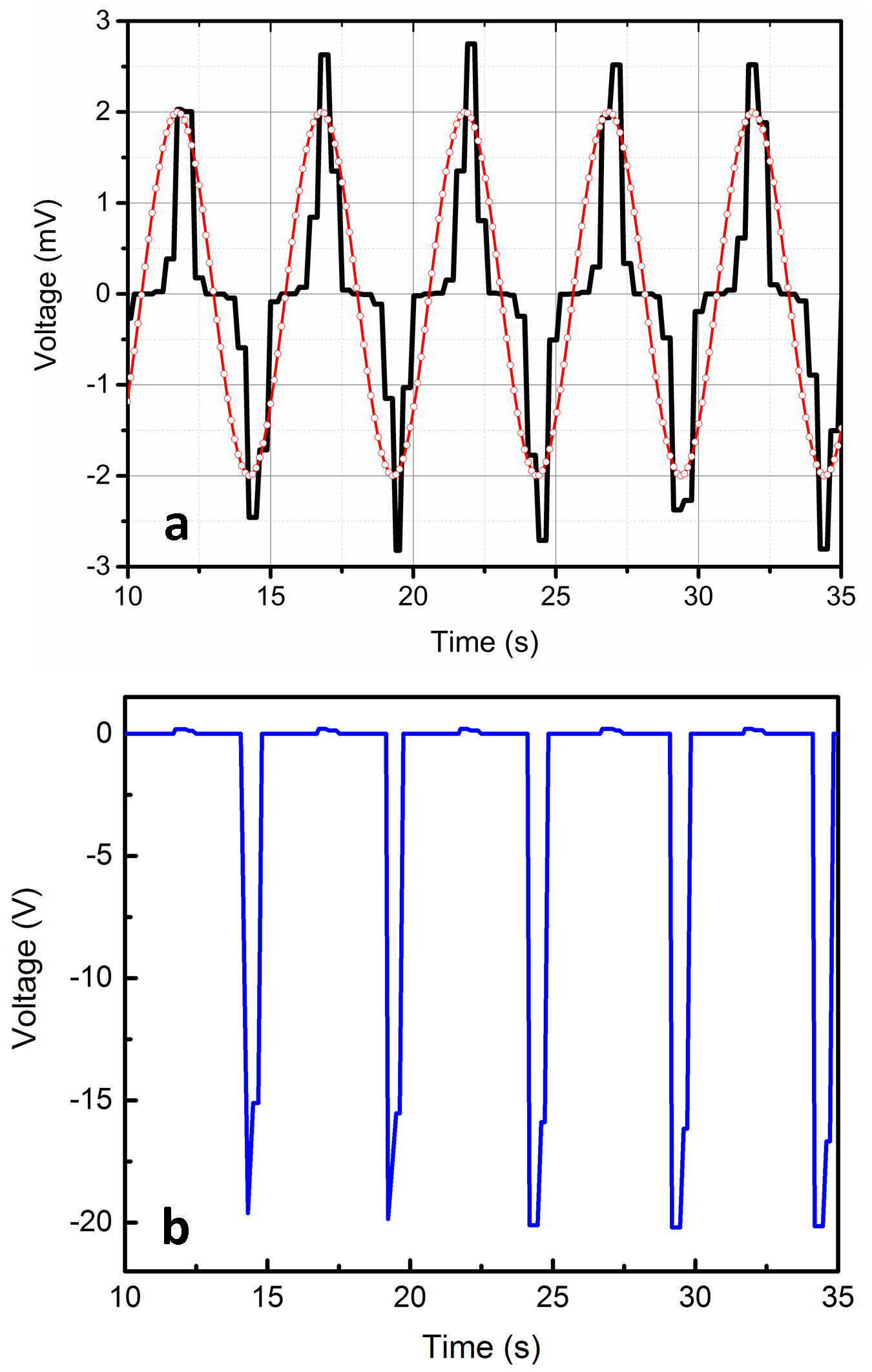}
\caption{Appearance of the diode effect in $Tl_{2}Ba_{2}CaCu_{2}O_{8}$ superconducting bridge
($T=100 K$ and $H=0$) at the increase of the bias current. The amplitude of sinusoidal AC current is $30 mA$ in panel (a) (shown by circles, arb. units) and $35 mA$ in panel (b); the frequency is $0.2Hz$ in both cases.}
\label{Figure2}
\end{figure}
This temperature was chosen for two reasons: 1) to stay closer to $T_{c}$ for keeping the critical current relatively low – at higher currents the transition to normal state could have irreversibly damaged the precious bridge; 2) at this temperature, as follows from Fig. 3, we have full resistive transition in zero fields. The microbridge was fabricated from a $Tl_{2}Ba_{2}CaCu_{2}O_{8}$ film with $%
T_{c}\approx 101 K$ (Fig. 3(a)). The resistive transition of the bridge is
shown in Fig. 3(b).
\begin{figure}
\includegraphics[width=0.9\linewidth]{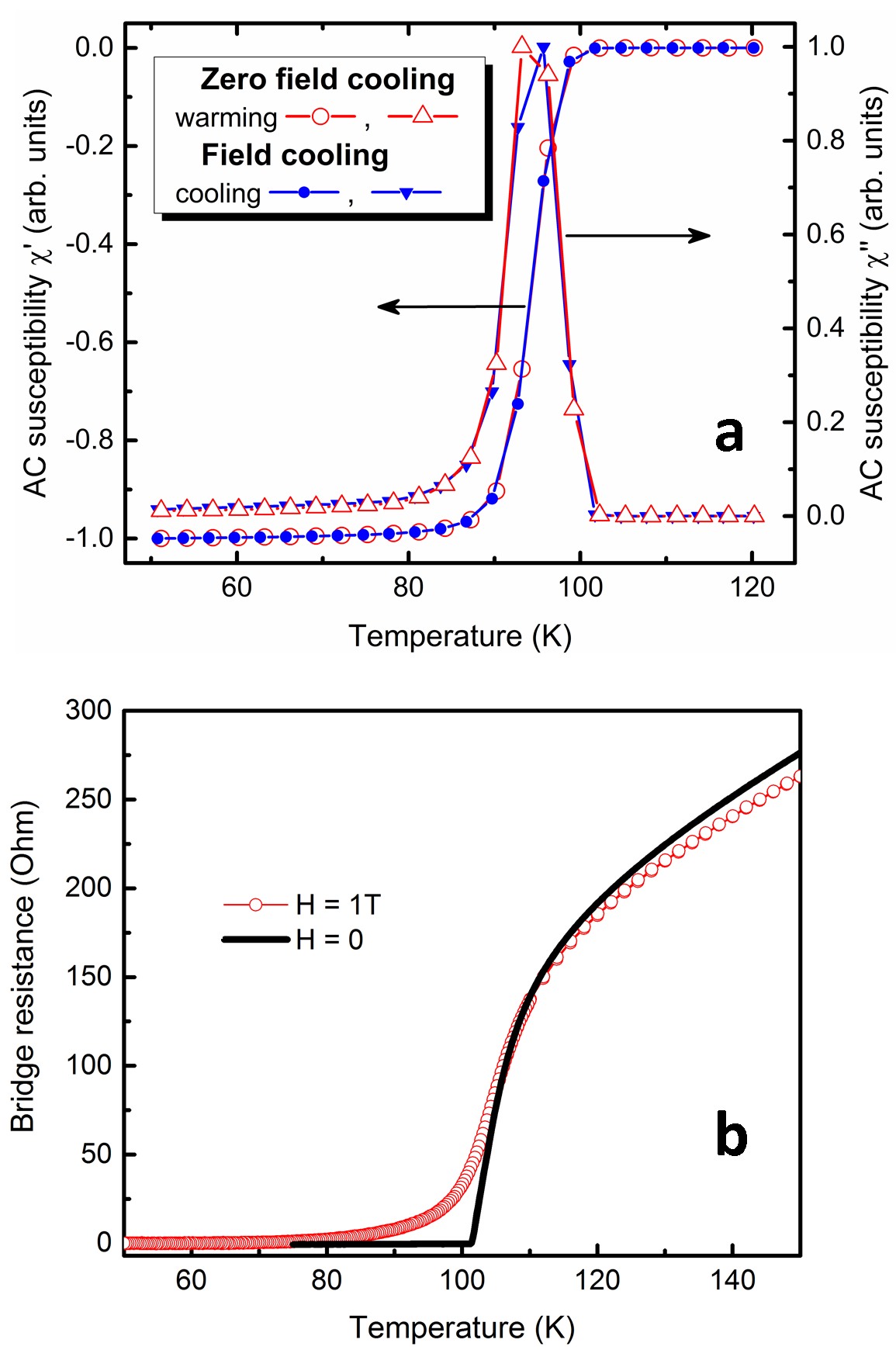}
\caption{Superconducting transition in $Tl_{2}Ba_{2}CaCu_{2}O_{8}$ film
recorded by AC susceptibility (panel (a); measurement frequency 400Hz, AC amplitude $5Oe$), and in lithographically processed bridge via resistivity (panel (b); measurement current $1\mu A$).}
\label{Figure3}
\end{figure}
The film thickness was approximately $1%
{\mu}%
m$, and the bridge dimensions were $20%
{\mu}%
m$ in width and $2mm$ in length. A defining aspect of these measurements is
that they were performed under strictly zero external magnetic field, $H=0$.
Measurements under $H=+100Oe$ and $H=-100Oe$ showed no observable difference
from the zero-field case (Fig. 4).
\begin{figure}
\includegraphics[width=0.9\linewidth]{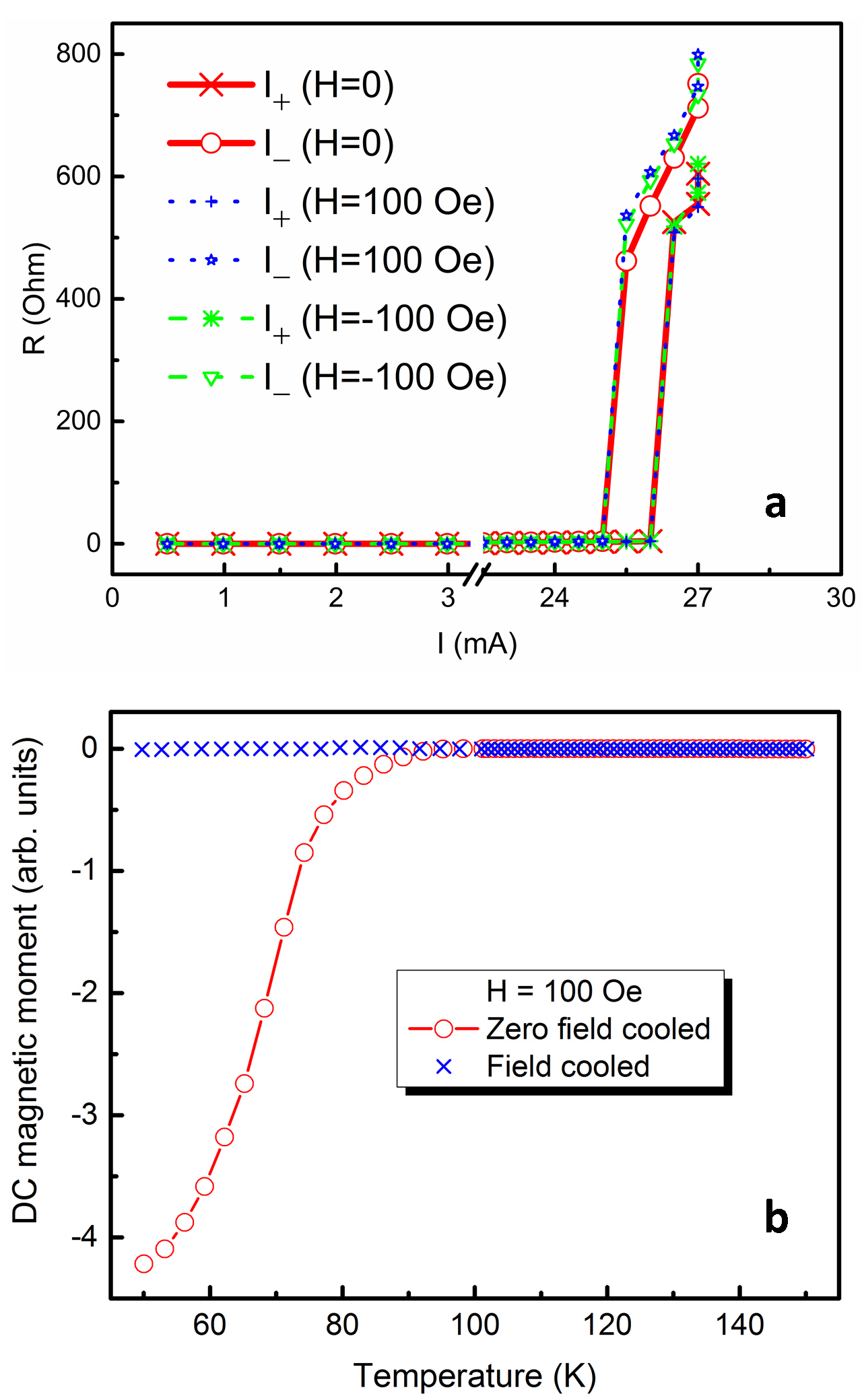}
\caption{Nonreciprocal superconducting transition in $%
Tl_{2}Ba_{2}CaCu_{2}O_{8}$ bridge is almost similar for $H=0$ and $H=\pm 100Oe$
(panel (a)). At the same time, as follows from panel (b), $100Oe$
exceeds the value of $H_{c1}$ of this material.}
\label{Figure4}
\end{figure}
These results therefore indicate that copper-oxide superconductors can
exhibit TRS breaking through internal mechanisms intrinsic to the
superconducting state.
\bigskip
\section{\label{sec:level1}Discussion}
The central experimental observation of this work is the emergence of a
pronounced superconducting diode effect in $Tl_{2}Ba_{2}CaCu_{2}O_{8}$
microbridges at $T=100 K$ under strictly zero applied magnetic field (Figs. 2
and 4). The nonreciprocal response remains unchanged within experimental
resolution under applied fields of $\pm 100Oe$, exceeding the nominal $%
H_{c1} $of the material.

Similar zero-field diode behavior has recently been reported in $%
Bi_{2}Sr_{2}CaCu_{2}O_{8+{\delta} }$ flakes \cite{ref7,ref9}. The present
results demonstrate that this phenomenon is not limited to a single
bismuth-cuprate family but also appears in a structurally related
thallium-based compound. Given the chemical and electronic similarities
between $Bi-$ and $Tl-$based cuprates, the consistency between these
independent observations strengthens the case that the effect reflects an
intrinsic property of the superconducting state rather than a
material-specific artifact.
\subsection{\label{sec:level2}Symmetry considerations}
Nonreciprocal superconducting transport generally requires the simultaneous
breaking of TRS and IS \cite%
{ref5}. In many experimental realizations of the superconducting diode
effect, TRS breaking is introduced by an external magnetic field \cite{ref1,ref2,ref3,ref4}. In contrast, the present measurements were conducted under
strictly zero applied field.

If the observed diode response is intrinsic, it implies that TRS breaking
may arise internally in the superconducting state. Such a possibility has
been discussed in the context of loop-current models of the pseudogap phase %
\cite{ref8,ref11} and in scenarios involving chiral or complex order
parameters \cite{ref12,ref13,ref14}. At the same time, we note that inversion
symmetry in our microbridge geometry may be broken extrinsically by
structural asymmetry at the film--substrate interface or by lithographic
processing. In that case, intrinsic TRS breaking combined with extrinsic IS
breaking would suffice to generate nonreciprocal response without invoking
inversion symmetry breaking in the bulk crystal.
\subsection{\label{sec:level2}Constraints on theoretical descriptions}
Rather than conclusively ruling out specific microscopic mechanisms, the
present results provide a symmetry-based constraint. Superconducting states
that strictly preserve TRS in equilibrium would require additional
mechanisms to explain zero-field nonreciprocity.

Preformed-pair and phase-fluctuation approaches \cite{ref15,ref16,ref17},
RVB-based models \cite{ref18,ref19,ref20}, and many density-wave descriptions
of the pseudogap \cite{ref21,ref22,ref23,ref24,ref25,ref26} do not, in their simplest
formulations, require spontaneous TRS breaking in the superconducting state.
If the diode effect observed here is intrinsic and reproducible across
cuprate families, these frameworks may require extension to incorporate
TRS-breaking components or nonequilibrium effects.

Conversely, models allowing spontaneous TRS breaking, including loop-current
states \cite{ref8,ref11}, or chiral superconducting order parameters \cite%
{ref12,ref13,ref14}, are naturally compatible with the symmetry requirements
of the diode effect. However, further experimental discrimination between
these possibilities remains necessary.
\subsection{\label{sec:level2}Nonequilibrium aspects}
An additional observation is that the diode effect appears above a finite
current threshold (Fig. 5).
\begin{figure}
\includegraphics[width=0.9\linewidth]{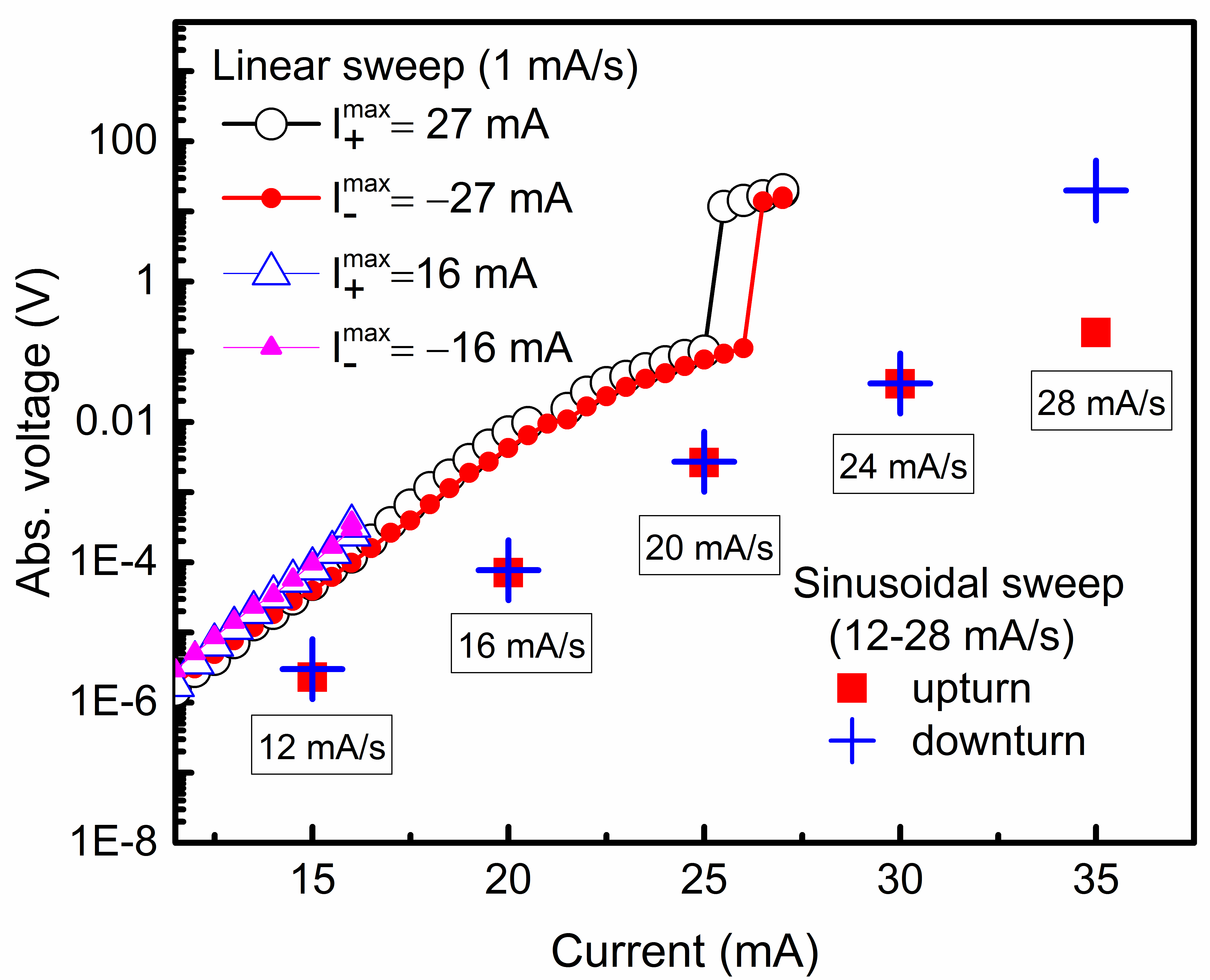}
\caption{Demonstration of possible nonequilibrium effects in the resistive
state of the bridge. The top datasets show absolute values of voltage measured
at linear current sweep to $16 mA$ and $27 mA$ with the relatively slow rate $1 mA/s$ (the lines are guide for eyes). These
curves (and many others measured) coincide at intersecting current values.
As soon as the current exceeds 25 mA, irreversibility appears. At the
sinusoidal bias current, the level of voltage is significantly lower (the bottom data points) because of much higher scanning speed (indicated on labels), and the irreversibility comes in at noticeably higher current amplitudes.}
\label{Figure5}
\end{figure}

This raises the possibility that nonequilibrium
superconducting dynamics may enhance or reveal symmetry-breaking tendencies
not prominent in the equilibrium state. Most microscopic theories of cuprate
superconductivity address equilibrium properties. If finite current density
plays a role in stabilizing or amplifying TRS-breaking components, the
observed diode behavior may reflect a current-driven phenomenon rather than
a purely equilibrium order parameter symmetry.

Further systematic studies, including current and temperature dependence, thickness scaling,
and complementary probes of TRS breaking, will be necessary to clarify this
distinction. The present measurements, together with recent reports in
Bi-based cuprates \cite{ref7,ref9}, suggest that zero-field
superconducting diode behavior may be a more general feature of
high-temperature superconductors than previously recognized.

While the data do not uniquely identify a microscopic mechanism, they impose
a robust symmetry constraint: any comprehensive theory of cuprate
superconductivity must accommodate the possibility of time-reversal symmetry
breaking, whether intrinsic to the equilibrium state or emerging under
finite current.

Continued experimental investigation across different cuprate families and
measurement geometries will be essential for determining the origin and
universality of this phenomenon.
\bigskip
\section{\label{sec:level1}Summary}
The above considerations lead to the following conclusion: a viable
theoretical model of HTSC in cuprates may need to incorporate TRS breaking.
This significantly narrows the set of compatible theoretical proposals,
although substantial uncertainty remains. Further constraints may arise from
examining additional signatures predicted by specific theories -- for
example, structural modifications detectable through X-ray or neutron
reflections. It is also possible that nonequilibrium electronic effects play a role in TRS breaking, as can be deduced from Fig.\ref{Figure5}. As follows from this figure, the diode effect emerges at
sufficiently high current densities. We should also refer to Fig. 2, which shows that the same bridge with the diode effect at the applied current amplitude $35 mA$  displays bipolar symmetry and markedly
lower resistance at the current amplitude smaller by $30\%$. Continued studies
of the diode effect in HTSCs may therefore yield further insight into the
underlying mechanism of superconductivity.
\begin{acknowledgments}
This research was supported in part by the ONR Grant No. N00014-24-1-2595.
We are grateful to the Physics Art Frontiers for the provided technical
assistance.
\end{acknowledgments}


\end{document}